\documentstyle[prb,12pt,aps,preprint,amssymb]{revtex}
\draft
\tighten
\def\varlatex{{\rm L\kern-.36em\raise.3ex\hbox{A}\kern-.15em
    T\kern-.1667em\lower.7ex\hbox{E}\kern-.125emX}}
\global\firstfigfalse
\global\firsttabfalse
\begin{document}
\language=0
\pagestyle{empty}
\input{psfig}
\def\bild#1#2#3#4#5#6{\centerline{\hbox{\psfig{figure=#1,height=#2
,bbllx=#3bp,bblly=#4bp,bburx=#5bp,bbury=#6bp,clip=}} }}
\def\ket#1{\vert\, #1 \, >}
\def\bra#1{< \, #1 \, }
\def\acct#1{#1^\prime }
\def\facu#1#2{\left( \begin{array}{c} #1 \\ #2 \end{array} \right)}
%
%
\def\slash#1{\setbox0=\hbox{$#1$}\setbox1=\hbox{$/$}\if\wd0>\wd1
\copy0\kern-\wd0\hbox to
\wd0{\hss\kern.1em\box1\hss}\else\copy1\kern-\wd1\hbox to
\wd1{\hss\box0\kern.1em\hss}\fi}
%
\title{ON PROTON AND DELTA WAVE FUNCTIONS \\ }
\vspace{-0.6 true cm}
\author{N. G. STEFANIS\thanks{Invited talk presented at the
Workshop on Exclusive
Reactions at high Momentum Transfer, Elba, Italy, 24-26 June, 1993, to
be published in the Proceedings.}
{\rm and} M. BERGMANN}
\address{Institut f\"ur Theoretische Physik II, \\
Ruhr-Universit\"at Bochum, D-44780 Bochum, Germany\\
E-mail: nicos@hadron.tp2.ruhr-uni-bochum.de}
\maketitle
\thispagestyle{empty}
\vspace{-0.5 true cm}
\begin{abstract}
Physical wave functions for the nucleon and the $\Delta^{+}$ isobar are
presented, which unify the best features of previous models. With these
wave functions we can calculate elastic
form factors and the decays of the charmonium levels
${}^{3}S_{1}$, ${}^{3}P_{1}$, ${}^{3}P_{2}$ into $p\bar p$
in agreement with the data.
A striking scaling behavior between
$R= \vert G_{\text{M}}^{\text{n}} \vert / G_{\text{M}}^{\text{p}}$
and the coefficient $B_{4}$ of the Appell
polynomial decomposition of the nucleon distribution amplitude is found;
the implications for elastic nucleon cross sections are discussed.
\end{abstract}
\vspace{-0.2 true cm}
\section{INTRODUCTION}
\vspace{-0.4 true cm}
An interesting testing ground for applications of perturbative QCD
has emerged in the study of exclusive processes and elastic form
factors of few-quark systems.
Such systems can be described within a convolution
formalism~\cite{LB80} assuming factorization of highly off-shell
or large transverse momentum regions of phase space from regions of low
momenta necessary to form bound states. Recent progress~\cite{LS92}
in Sudakov-suppression techniques provides support for the conjectured
infrared protection of the perturbative picture.

For modelling of the nucleon and its low resonances, elastic form
factors play a key role because they provide an integrated view of the
implications of QCD from low to high $Q^{2}$. Thus they offer a powerful
link between theoretical concepts and measurements and they can serve to
test both the scaling properties as well as the detailed structure of
the nucleon wave function.
The theoretical tools for such a description are provided by the
hard-scattering amplitude which describes the perturbative quark-gluon
interaction in a particular process, and the probability amplitude for
finding the three-quark valence state in the scattered nucleon or
nucleon resonance:
$
 \Phi_{\text{N}}(x_{\text{i}},Q^{2}).
$
A major theme of this talk will be to examine how QCD deals with the
derivation of such distribution amplitudes for the nucleon and the
$\Delta^{+}$ isobar, focusing our attention on recent developments.
\vspace{-0.6 true cm}
\section{GENERAL FEATURES}
\vspace{-0.4 true cm}
The momentum-scale dependence of
$
 \Phi_{\text{N}}(x_{\text{i}},Q^{2})
$
is given by
\begin{equation}
  \Phi_{\text{N}}(x_{\text{i}},Q^{2})=
  \Phi_{\text{as}}(x_{\text{i}})\sum_{n=0}^{\infty}B_{\text{n}}
  \tilde \Phi_{\text{n}}(x_{\text{i}})\biggl (
  \frac{\alpha_{\text{s}}(Q^{2})}
  {\alpha_{\text{s}}(\mu^{2})}\biggr )^{\gamma_{\text{n}}},
\end{equation}
in which $\{\Phi_{\text{n}}\}_{0}^{\infty}$ are orthonormalized
eigenfunctions of the interaction kernel of the evolution
equation~\cite{LB80} expressed in a truncated basis of Appell
polynomials of maximum degree $M$, and
$\Phi_{\text{as}}(x_{\text{i}})=120x_{1}x_{2}x_{3}$
is the asymptotic form of the nucleon distribution amplitude.
The corresponding eigenvalues $\gamma_{\text{n}}$ turn out~\cite{Pes79}
to be the anomalous dimensions of multiplicatively renormalizable
$I_{1/2}$ baryonic operators of twist three.
Because the $\gamma_{\text{n}}$ are positive fractional numbers
increasing with $n$, higher terms in this expansion are gradually
suppressed. A basis including a total of $54$ eigenfunctions ($M=9$)
together with the associated normalization coefficients and anomalous
dimensions is given in Ref.\ \onlinecite{BS93c}.

The derivation of the nucleon distribution amplitude from QCD is
intimately connected with confinement and employs nonperturbative
methods.
Using the properties of the Appell polynomials, the inverse of Eq.~(1)
determines the (nonperturbative) expansion coefficients $B_{\text{n}}$:
\begin{equation}
  B_{\text{n}}(\mu^{2})=\frac{N_{\text{n}}}
  {120}\int_{0}^{1}[dx]\tilde
  \Phi_{\text{n}}(x_{\text{i}})\Phi_{\text{N}}(x_{\text{i}}, \mu^{2}),
\end{equation}
so that the ``renormalization-group improved'' coefficients
$
 B_{\text{n}}(Q^{2})
$
are given by
\begin{equation}
  B_{\text{n}}(Q^{2}) = B_{\text{n}}(\mu^{2})
  \text{exp}
  \Biggl\{ -\int_ { \alpha_ {\text{s}} (\mu^{2})
                  } ^
                  { \alpha_ {\text{s}} (Q^{2})
                  }
  \frac {d\alpha} {\beta(\alpha )}
  \gamma_{\text{n}}(\alpha )\Biggr\}
  \approx
  B_ {\text{n}} (\mu^{2})
     \Biggl \{
             \frac {\ln (Q^{2}/\Lambda_{\text{QCD}}^{2})
             }
             {\ln (\mu^{2}/\Lambda_{\text{QCD}}^{2})
             }
     \Biggr \} ^{-\gamma_{\text{n}}}.
\end{equation}
In terms of the moments of the nucleon distribution amplitude,
\begin{equation}
  \Phi_{\text{N}}^{(\text{i}0\text{j})}(\mu^{2}) =
  \int_{0}^{1}[dx]x_{1}^{\text{i}}x_{2}^{0}x_{3}^{\text{j}}
  \Phi_{\text{N}}(x_{i},\mu^{2}),
\end{equation}
Eq.~(3) becomes
\begin{equation}
  \frac{B_ {\text{n}} (\mu^{2})}{\sqrt{N_{\text{n}}}} =
  \frac{\sqrt{N_{\text{n}}}}{120}
  \sum_{\text{i,j}=0}^{\infty}a_{\text{ij}}^{\text{n}}\
  \Phi_{\text{N}}^{(\text{i}0\text{j})}(\mu^{2}),
\end{equation}
where the projection coefficients
$
 a_{\text{ij}}^{\text{n}}
$
are calculable to any order $M$. Specifically, those up to order $M=9$
have been tabulated in Refs.\ \onlinecite{BS93c,Ber93}.

To determine the moments,
a short-distance operator product expansion is performed at some
spacelike momentum $\mu^{2}$ where quark-hadron duality is
valid.~\cite{CZ84b}
One considers matrix elements of appropriate three-quark operators
which are related to moments of the covariant distribution
amplitudes~\cite{HKM75} $V$, $A$, and $T$:
$
 \Phi_{\text{N}}(x_{\text{i}})=V(x_{\text{i}})-A(x_{\text{i}}),
$
$
 \Phi_{\text{N}}(1,3,2)+\Phi_{\text{N}}(2,3,1)=2T(1,2,3)
$
with
V(1,2,3)=V(2,1,3), A(1,2,3)=-A(2,1,3), and T(1,2,3)=T(2,1,3).
\vspace{-0.6 true cm}
\section{DISTRIBUTION AMPLITUDES OF THE NUCLEON AND THE
$\Delta^{+}$ ISOBAR}
\vspace{-0.4 true cm}
Based on QCD sum-rule calculations, useful theoretical constraints on
the moments of baryon distribution amplitudes have been
obtained.~\cite{CZ84b,COZ89a,KS87,CP88,Far88}
Physical wave functions and observables for the
nucleon~\cite{CZ84b,COZ89a,KS87} and the $\Delta^{+}$
isobar~\cite{CP88,Far88} are then calculated using the full set of these
constraints to determine the first few expansion coefficients
$B_{\text{n}}$ in a truncated basis of Appell polynomials.
Depending on the value of
$\Lambda_{\text{QCD}}$,
these models predict approximately the right
size and $Q^{2}$-evolution of $G_{\text{M}}^{\text{p}}$,
while they give
$R = \vert G_{M}^{n}\vert /G_{M}^{p}\le 0.5$.
An alternative nucleon distribution amplitude was proposed~\cite{GS86}
to give
$\vert G_{M}^{n}\vert \ll G_{M}^{p}$,
in accordance to phenomenological data analyses~\cite{GK85}
and the latest high-$Q^{2}$ SLAC
data~\cite{Arn86} at the expense that some of the amplitude moments
cannot match the sum-rule requirements~\cite{CZ84b} in the allowed
saturation range.~\cite{Ste89}

However, several crucial
questions have to be resolved: For instance, does the {\it optimum}
solution to the sum rules automatically yield {\it best} agreement
with the data? Do solutions exist with characteristics {\it distinctive}
from those of the COZ and the GS amplitudes? If so, what are the
fundamental ordering parameters to classify these solutions?
In recent works~\cite{SB92a,SB92b} we have shown that it is indeed
possible to amalgamate the best features of COZ-type~\cite{COZ89a} and
GS-type~\cite{GS86} nucleon distribution amplitudes into a hybrid-like
amplitude, we termed the ``heterotic'' solution (see Fig.~1).

In order to develop a credible
nucleon distribution amplitude, we employ a $\chi^{2}$ criterion which
parametrizes the deviations from the sum-rule intervals according to
the moment order. This ``hierarchical'' treatment of the sum rules takes
into account the higher stability of the lower-level moments~\cite{Ste89}
and does not overestimate the significance of the still unverified
constraints~\cite{COZ89a} for the third-order moments. [For more
details, see Ref.\ \onlinecite{BS93a}.]

The underlying assumption
is that contributions of higher-order terms
are either negligible or of minor importance relative to those of
second-order. Then the model space is also truncated at states with
bilinear
correlations of fractional momenta and the pattern of solutions found in
this order should dominate the (orthonormalized) Appell polynomial
series
at every order of truncation. In this way the parameter space of the
Appell decomposition coefficients can be systematically scanned
seeking for local minima of $\chi^{2}$.
Using for the first and second order moments either the COZ or the KS
sum-rule constraints in conjunction with those of COZ for the
third-order moments, a simple {\it scaling relation} between the ratio
$R$ and the expansion coefficient $B_{4}$ emerges as one progresses
through the generated solutions.~\cite{BS93b}

We have plotted in the ($B_{4},R$) plane interpolating solutions to the
COZ sum rules ($+$ labels) and such to a combined set of KS/COZ
sum rules ($\circ$ labels).
As it turns out (Fig.~2), there is no significant difference
between the two treatments and this insensitivity justifies
the whole approach. The presented curves are fits to the local minima of
the COZ sum rules (solid line) and the KS/COZ sum rules (dotted line).
They constitute an {\it orbit} with respect to $\chi^{2}$, beginning in
the heterotic region (small $R$ and large positive $B_{4}$) and
terminating past the COZ cluster (large $R$ and large negative $B_{4}$).


The lower part of the orbit is associated with the heterotic solution
which corresponds to the smallest possible ratio still compatible with
the sum-rule constraints.
The upper region of the orbit controls COZ-type amplitudes and contains
a cluster of solutions densely populating the orbit in the $R$-interval
$0.455 \div 0.495$ (see the inset in Fig.~2).
This cluster contains the amplitudes $COZ^{\text{opt}}$,
$KS/COZ^{\text{opt}}$ which are associated with the absolute minima
of $\chi^{2}$ and play the role of strange attractors for all other
solutions with similar features.~\cite{BS93a}
GS-type amplitudes correspond to local minima of $\chi^{2}$ at
considerably lower levels of accuracy and thus they constitute
in the ($B_{4},R$) plane an isolated region (an ``island'') that is
separated from the characteristic orbit by a large $\chi^{2}$ barrier.
The profiles of the distribution amplitudes across the orbit change in
an orderly sequence of gradations with some mixture of COZ and
GS characteristics until the COZ amplitude is transmuted into the
heterotic solution.~\cite{BS93a}

\begin{table}
\caption{Theoretical parameters defining the nucleon distribution
         amplitudes discussed in the text. The "hybridity" angle
         $\vartheta$ is discussed in [19].
}
\squeezetable
\begin{tabular}{lrrrrrrrrc}
 Model       & $B_{1}$\ \ &  $B_{2}$\ \ &   $B_{3}$\ \  &    $B_{4}$\ \ &
$B_{5}$\ \ &
$\vartheta$[deg]\  &  R\ \ & $\chi^{2}$\ & Symbol \cr
\tableline
 $Het     $ & 3.4437 &  1.5710 &   4.5937  &   29.3125 &    -0.1250  &    -1.89
 &    .104 &  33.48  &{\Large $\bullet$} \cr
 $Het^\prime $& 4.3025&  1.5920 &   1.9675  &  -19.6580 &     3.3531  &
24.44  &    .448 &  30.63 &{\Large $\bullet$} \cr
 $COZ^{\text{opt}}$ & 3.5268 &  1.4000 &   2.8736  &   -4.5227 &     0.8002  &
   9.13  &    .465 &   4.49 &$\blacksquare$    \cr
 $COZ^{\text{up}}$ & 3.2185 &  1.4562 &   2.8300  &  -17.3400 &     0.4700  &
5.83  &    .4881& 21.29 &$+$ \cr
 $COZ      $ & 3.6750 &  1.4840 &   2.8980  &   -6.6150 &     1.0260  &
10.16  &    .474 &  24.64 &$\Box$   \cr
 $CZ       $ & 4.3050 &  1.9250 &   2.2470  &   -3.4650 &     0.0180  &
13.40  &    .487 & 250.07 &$\blacklozenge$ \cr
 $KS^{\text{low}}$ & 3.5818 &  1.4702 &   4.8831  &   31.9906 &     0.4313  &
 -0.93  &    .0675&  36.27  &$\circ$ \cr
 $KS/COZ^{\text{opt}}$  & 3.4242 &  1.3644 &   3.0844  &   -3.2656 &     1.2750
 &     9.47  &    .453 &   5.66 &$\circ$    \cr
 $KS^{\text{up}}$ & 3.5935 &  1.4184 &   2.7864  &  -13.3802 &   2.0594   &
13.82  &    .482 & 40.38 &$\circ$ \cr
 $KS       $ & 3.2550 &  1.2950 &   3.9690  &    0.9450 &     1.0260  &
2.47  &    .412 & 116.35 &$\Diamond$ \cr
 $GS^{\text{opt}} $ & 3.9501 &  1.5273 &  -4.8174  &    3.4435 &     8.7534  &
  80.87  &    .095 &  54.95 &$\blacktriangle$ \cr
 $GS^{\text{min}} $ & 3.9258 &  1.4598 &  -4.6816  &    1.1898 &     8.0123  &
  80.19  &    .035 &  54.11 &$\blacktriangledown$ \cr
 $GS       $ & 4.1045 &  2.0605 &  -4.7173  &    5.0202 &     9.3014  &
78.87  &    .097 & 270.82 &$\bigtriangleup$ \cr
\end{tabular}
\end{table}

Let us now turn to models with functional representations
which make use of higher Appell polynomials in connection with
additional {\it ad-hoc} cutoff-parameters.~\cite{Sch89,HEG92}
The inset in Fig.~2 shows how such models~\cite{Sch89} (stars)
and~\cite{HEG92} (light upside-down triangles)
group around the optimum amplitudes $COZ^{\text{opt}}$ and
$KS/COZ^{\text{opt}}$, thus establishing the scaling relation between
$R$ and $B_{4}$ in a much more general context.
This result suggests that the inclusion of higher-order Appell
polynomials in the nucleon distribution amplitude is a marginal effect,
as conjectured above. Those model amplitudes~\cite{Sch89,HEG92}
which appear as isolated points scattered towards the GS island are
unacceptable on physical grounds, either because they exhibit unrealistic
large oscillations in the longitudinal momentum fractions~\cite{Sch89} or
because they yield a wrong evolution behavior for the nucleon form
factors.~\cite{HEG92}

One place to test these results is in the data for the elastic cross
sections
$\sigma_{\text{p}}$ and $\sigma_{\text{n}}$.
For small scattering angles, where the terms
$\propto tan^{2}(\theta/2)$
can be neglected, there are two main possibilities for the ratio
$
 \sigma_{\text{n}}/ \sigma_{\text{p}}.
$
If the Dirac form factor $F_{1}^{\text{n}}$ is zero or small compared
to the Pauli form factor $F_{2}^{\text{n}}$,~\cite{GK85} then
$\sigma_{\text{n}}$ should be due only to the higher-order term
$F_{2}^{\text{n}}$. At large $Q^{2}$ the ratio would become
($M_{\text{N}}$ is the nucleon mass)
$
 \frac {\sigma_{\text{n}}} {\sigma_{\text{p}}} \Rightarrow
 \Bigl( \frac {
              C_{2}^{\text{n}}
                              } {
                                 C_{1}^{\text{p}}
                                }  \Bigr)^{2}
                                    \frac {1}{
                                              4M_{\text{N}}^{2} Q^{2}
                                              }
$
and would decrease with increasing $Q^{2}$ due to the extra power of
$1/Q^{2}$ of the Pauli form factor. Alternatively, if
$F_{1}^{\text{n}}$
is comparable to $F_{2}^{\text{n}}$, then $\sigma_{\text{n}}$ would
eventually be due to $F_{1}^{\text{n}}$ at large $Q^{2}$. Then the ratio
$
 \sigma_{\text{n}}/ \sigma_{\text{p}}.
$
would be given by some constant determined by the nucleon wave functions
$
 \frac {\sigma_{\text{n}}} {\sigma_{\text{p}}} \Rightarrow
 \Bigl( \frac {
              C_{1}^{\text{n}}
                              } {
                                 C_{1}^{\text{p}}
                                }  \Bigr)^{2}.
$
In these expressions, the wave-function characteristics
are parametrized by the (dimensionful) coefficients
$C_{\text{i}}$, which are functions of the expansion coefficients
$B_{\text{n}}$ and the ``proton decay constant''
$
 \vert f_{\text{N}}\vert = (5.0\pm 0.3)\times 10^{-3} GeV^{2}.
$

The principal result from the above discussion is that in the
intermediate $Q^{2}$ domain,
$
 \sigma_{\text{n}}/ \sigma_{\text{p}}
$
should be within the range $0.238$ and $0.01$.
Comparing with available data~\cite{Roc92}, we see that the measured
$\sigma_{\text{n}}/\sigma_{\text{p}}$
enters the estimated range already at
$
 Q^{2}\approx 8 GeV^{2}/c^{2}
$
(see Fig.~3).


In view of these results, it is worth remarking that the present
accuracy of QCD sum rules seems to be sufficient to limit
$
 \sigma_{\text{n}}/\sigma_{\text{p}}
$
within the observed region.
Fig.~3 shows that the available data in the range
$
 Q^{2}\approx (8 \div 10) GeV^{2}/c^{2}
$
are well below the calculated upper bound and still decreasing.
This indicates that distribution amplitudes which give
$
 \vert G_{\text{M}}^{\text{n}}\vert / G_{\text{M}}^{\text{p}}
 \approx 0.5
$
may be in contradiction to experiment because they yield
a Dirac form factor $F_{1}^{\text{n}}$ which starts to overestimate
the data already at
$
 Q^{2}\approx 8 GeV^{2}/c^{2}.
$
On the contrary, models which give a small value of
$
 \vert G_{\text{M}}^{\text{n}}\vert / G_{\text{M}}^{\text{p}}
$
can explain the data only under the assumption that in this $Q^{2}$
region the Pauli contribution is still dominant. Fig.~4 serves not only
to amplify the preceeding discussion but also to advertise the
consistency of the heterotic model with the form factor data. A similar
good agreement with the data is found also for the axial form
factors.~\cite{SB92a,Ste92}


We now turn our attention to the $\Delta^{+}$ isobar.
It was pointed out~\cite{CGS87} that model amplitudes for the nucleon
are characterized by an anticorrelation pattern between
$ G_{\text{M}}^{\text{n}}$ and $G_{\text{M}}^{*}$. COZ-like models yield
$\vert G_{\text{M}}^{\text{n}}\vert / G_{\text{M}}^{\text{p}}
\leq 0.5$ and
$
 \vert G_{\text{M}}^{*}\vert / G_{\text{M}}^{\text{p}}
$
small, while GS-like models lead to the reverse situation.
This pattern was derived, under the assumption that
the $\Delta$ amplitude can be crudely modelled by the symmetric part of
the nucleon distribution amplitude. Fig.~5 shows that the transition
form factor calculated with the nucleon heterotic amplitude and more
realistic $\Delta$ amplitudes, derived from QCD sum
rules,~\cite{CP88,Far88} is positive with a magnitude between those of
previous models.
In order to obtain an optimum distribution amplitude
for the
$\Delta$, we try to comply with the constraints of
the CP~\cite{CP88} and FZOZ~\cite{Far88} analyses
simultaneously.~\cite{SB92b} This concept leads to a hybrid-like
amplitude, denoted again ``heterotic''. This solution fulfills all FZOZ
constraints and provides the best possible compliance with the CP
constraints. In addition, it gives the best agreement
with the data (see Fig.~5 and Ref.\ \onlinecite{Stu93}).
In particular, when including the effect of perturbative (i.e., {\it
logarithmic}) $Q^{2}$ evolution of the expansion coefficients
$B_{\text{n}}$, the combined use of the heterotic amplitudes for the
nucleon and the $\Delta^{+}$ yields a form factor behavior which
conforms
with the observed decrease of available data within their quoted
errors.~\cite{SB92b}


There is yet another type of solution for the $\Delta$
amplitude---compatible with the sum rules~\cite{CP88,Far88}---but in
sizeable disagreement with the data. This solution ($FZOZ^{\text{opt}}$)
is obtained by demanding that $G_{\text{M}}^{*}$ calculated with
$COZ^{\text{opt}}$ is positive.
Thus, as in the nucleon case, optimum agreement with the (existing) sum
rules {\it does not} automatically entail best agreement with the data.

Exclusive decays of charmonium levels to $p\bar p$ are very
sensitive to the nucleon distribution amplitude.
The branching ratio for the
decay of the $\chi_{\text{c}1}$ state ($J^{\text{CP}}=1^{++}$) into
$p\bar p$ is proportional to the decay amplitude $M_{1}$, which involves
$\Phi_{N}$ and $f_{\text{N}}$.
Inputing the heterotic amplitude,
$M_{1}$ is computed using an elaborated integration routine which
accounts for contributions near singularities.~\cite{Ber93} Thereby
we find $M_{1}^{\text{het}}=99849.6$ and as a result
$
 BR(^3P_{1}\to p\bar p/^3P_{1}\to \text{all})=0.77\times 10^{-2}\%
$,
which is in excellent agreement with the recent high-precision
experimental value~\cite{Arm92}
$(0.78\pm 0.10\pm 0.11)\times 10^{-2}$
of the E760 Colaboration at FNAL.

Analogously
for the $\chi_{\text{c}2}$ state ($J^{\text{PC}}=2^{++}$), we find
$M_{2}^{\text{het}}=515491.2$.
Setting~\cite{BGR81} $\alpha_{\text{s}}(m_{\text{c}})=0.210\pm 0.028$,
we then obtain
$
 BR({{}^3P_{2}\to p\bar p}/{{}^3P_{2}\to\text{all}})=
 0.89\times 10^{-2}\%
$
in excellent agreement with the FNAL value~\cite{Arm92}
$(0.91\pm 0.08\pm 0.14)\times 10^{-2}\%$.

Similar considerations apply also to the charmonium decay of the level
${}^{3}S_{1}$ with $J^{\text{PC}}=1^{--}$.
The partial width of $J/\psi$ (or $\chi_{\text{c}0}$) into $p\bar p$ is
$
  \Gamma({}^3S_{1}\to p\bar p) =
  (\pi\alpha_{\text{s}})^{6}{{1280}\over{243\pi}}
  {{{\vert f_{\psi}\vert}^{2}}\over{\bar M}}
  {\Bigg \vert {{f_{\text{N}}}\over{{\bar M}^{2}}}
  \Bigg \vert}^{4}{M_{0}^{2}},
$
where $f_{\psi}$ determines the value of the ${}^3S_{1}$-state wave
function at the origin. Its value can be extracted from the leptonic
width
$\Gamma({}^3S_{1}\to e^{+}e^{-}) = (5.36\pm 0.29) keV$~\cite{PDG92}
via the Van Royen-Weisskopf formula. The result
is ($m_{J/\psi}=3096.93 MeV$)
$\vert f_{\psi}\vert =409 MeV$.
The heterotic amplitude gives for this transition $M_{0}=13726.8$.
Then, using the previous parameters, it follows that
$\Gamma({}^3S_{1}\to p\bar p) = 0.14 keV$.
{}From experiment~\cite{PDG92}
it is known that $\Gamma(p\bar p)/\Gamma_{\text{tot}}=
(2.16\pm 0.11)\times 10^{-3}$ with $\Gamma_{\text{tot}}=(68\pm 10) keV$,
so that
$\Gamma({}^3S_{1}\to p\bar p) = 0.15 keV$ in remarkable agreement with
the model prediction.
The corresponding branching ratio is
$
 BR({{}^3S_{1}\to p\bar p}/{{}^3S_{1}\to\text{all}})=
 1.62\times 10^{-3}
$
with
$
 \Gamma_{\text{tot}}=(85.5{{+6.1}\atop {-5.8}})keV.
$
To effect the quality of these predictions, we quote the results
for the COZ amplitude:~\cite{COZ89b} [Note that these authors use the
rather arbitrary value $\alpha_{\text{s}}=0.3$.]
$
 BR(^3P_{1}\to p\bar p/^3P_{1}\to \text{all})=0.50\times 10^{-2}\%
$,
$
 BR({{}^3P_{2}\to p\bar p}/{{}^3P_{2}\to\text{all}})=
 1.6\times 10^{-2}\%
$, and
$\Gamma({}^3S_{1}\to p\bar p) = 0.34 keV$.
\vspace{-0.6 true cm}
\section{SUMMARY AND CONCLUSIONS}
\vspace{-0.4 true cm}
Given the apparent success of the heterotic
model~\cite{SB92a,SB92b,Ste92} in predicting a variety of observables
such as magnetic and transition form factors and several branching
ratios of exclusive decays of charmonium into $p\bar p$, it is optimistic
to believe that this approach---albeit approximative for a complete
analytical understanding of the nucleon distribution amplitude---is
sufficient of reproducing the observed phenomena. Since higher than order
$M=3$ expansion coefficients are unspecified by the present knowledge of
QCD sum rules, the model does not depend on unconstrained (higher-order)
parameters.
While higher-order effects on the nucleon distribution amplitude itself
are found to be large,~\cite{Sch89,HEG92} the agreement with the data is
actually not improved.~\cite{BS93c} This is also true for the optimized
version of the COZ amplitude~\cite{BS93d}, we have derived, which
represents the global minimum of $\chi^{2}$. We emphasize that the
(normalized) coefficients $B_{\text{n}}$ calculated via the
central values of the $10$ independent sum rules of
Ref.\ \onlinecite{COZ89a} {\it do not} correspond to a solution with
$\chi^{2}=0$. Although such a solution {\it must} exist, its
determination is not a trivial task. Furthermore, at relatively large
distances probed in present experiments, still uncalculable contributions
of higher twists are presumably more significant than higher-order terms
of the Appell polynomial series.  \par

{\it This work was supported in part by the Deutsche
Forschungsgemeinschaft and the COSY-J\"ulich project. \par} \par
\vspace{0.2 true cm}

\end{document}